\documentclass[aps,prl,preprint,groupedaddress]{revtex4-1}
\usepackage{graphicx}
\usepackage[latin1]{inputenc}
\usepackage{graphicx}
\usepackage{amssymb}
\usepackage{color}
\usepackage{float}
\usepackage{amsmath}
\usepackage{amsfonts}
\usepackage{dcolumn}
\usepackage{hyperref}
\usepackage{amsthm}
\usepackage{color}
\usepackage{bm}

\def\be{\begin{equation}}
\def\ee{\end{equation}}
\def\beq{\begin{eqnarray}}
\def\eeq{\end{eqnarray}}
\def\n{\nonumber}

\begin{document}

%\preprint{}

\title{Static trace free Einstein equations and stellar distributions}



\author{Sudan Hansraj} \email[]{hansrajs@ukzn.ac.za}
\affiliation{ Astrophysics and Cosmology Research Unit, School of Mathematics, Statistics and Computer Science,
University of KwaZulu-Natal, Private Bag X54001, Durban 4000, South Africa}

\author{Rituparno Goswami} \email[]{vitasta9@gmail.com}
\affiliation{ Astrophysics and Cosmology Research Unit, School of Mathematics, Statistics and Computer Science,
University of KwaZulu-Natal, Private Bag X54001, Durban 4000, South Africa}

\author{George Ellis} \email{gfrellis@gmail.com}
\affiliation{Department of Mathematics and Applied Mathematics, University of Cape Town, Private Bax X1, Rondebosch, 7701, South Africa}

\author{Njabulo Mkhize} \email{mkhizen18@gmail.com}
\affiliation{ Astrophysics and Cosmology Research Unit, School of Mathematics, Statistics and Computer Science,
University of KwaZulu-Natal, Private Bag X54001, Durban 4000, South Africa}

%\author{Sudan Hansraj}
%\author{Rituparno Goswami}
%\author{George Ellis}
%\author{Njabulo Mkhize}
%\affiliation{Astrophysics and Cosmology Research Unit, University of KwaZulu Natal}
%\email[]{hansrajs@ukzn.ac.za}
%\email[]{vitasta9@gmail.com}
%\email[]{gfrellis@gmail.com\hat{}}
%\email[] {mkhizen18@gmail.com}
%\thanks{}

\date{\today}

\begin{abstract}
We construct models of static spherical distributions of perfect fluid in  trace--free Einstein gravity theory. The equations governing the gravitational field are equivalent to the standard Einstein's equations however, their presentation is manifestly different which motivates the question whether new information would emerge due to the nonlinearity of the field equations. The incompressible fluid assumption does not lead to the well known Schwarzschild interior metric of Einstein gravity and a term denoting the presence of a cosmological constant is present on account of the integration process. The Schwarzschild interior is regained as a special case of a richer geometry. On the other hand, when the Schwarzschild geometry is prescribed, a constant density fluid emerges consistent with the standard equations.   A complete model of an isothermal fluid sphere with pressure and density obeying the inverse square law  is obtained. Corrections to the model previously presented in the literature by Saslaw {\it {et al}} are exhibited. The isothermal ansatz does not yield a constant gravitational potential in general but both potentials are position dependent. Conversely, it is shown that assuming a constant $g_{rr}$ gravitational potential does not yield an isothermal fluid in general as is the case in standard general relativity. The results of the standard Einstein equations are special cases of the models reported here. Noteworthy is the fact that whereas the previously reported isothermal solution was only of cosmological interest, the solution reported herein admit compact objects by virtue of the fact that a pressure-free hypersurface exists.  Finally we analyze the consequences of selecting the Finch--Skea metric as the seed solution. The density profiles match however there is a deviation between the pressure profiles with the Einstein case although the qualitative behaviour is the same. It is shown in detail that the model satisfies elementary requirements for physical plausibility such as a positive density and pressure, existence of a hypersurface of vanishing pressure, a subluminal sound speed, satisfaction of the weak, strong and dominant energy conditions as well as the Buchdahl mass--radius compactness requirement.

\end{abstract}

\pacs{}
%\keywords{}

\maketitle


\section{Introduction}

Despite the overwhelming successes of general relativity, the question of what the true theory of gravity is, is still  speculated on. This is so because some fundamental problems cannot be resolved within the framework of general relativity and ordinary matter. Specifically, the late time accelerated expansion of the universe as demonstrated by Supernova observations, matter surveys, and WMAP surveys, does not follow directly from general relativity in conjunction with ordinary matter. In attempting to address the problem an appeal has been made to exotic matter such as dark energy, or the cosmological constant first proposed by Einstein. The difficulty  is that there is a discrepancy of many orders of magnitude between the value  of the vacuum energy predicted by quantum field theory, and the value of the cosmological constant  deduced from astronomical observations of the expanding universe, if General Relativity is the correct theory of
gravity. The problem might be resolved if some other theory applied.  Therefore, questions about the fundamental theory abound.\\

One extension of the theory of general relativity is Lovelock gravity. Recently Dadhich \cite{NKD} argued that the pure Lovelock field equations constitute the correct gravitational equations. It is noteworthy that strong support for this claim is that  the first order of the Lovelock lagrangian corresponds identically to Einstein gravity so all the gains of general relativity are not lost. Moreover to second order, the result is the Gauss--Bonnet lagrangian which occurs in heterotic string theory \cite{gross}.  The Lovelock polynomial consists of terms quadratic in the Riemann tensor, Ricci tensor and Ricci scalar, however it is free of derivatives of these tensors. Remarkably all terms with derivatives of order more than 2 cancel out and the equations of motion are second order as expected for a theory of gravity. Competing ideas such as $f(R)$ theory \cite{starobinksi} must contend with ghost terms in the form of fourth order derivatives.\\

A simpler extension than either of these is using the trace-free Einstein equations  \cite{ellis1} to explain the differences in the values of the vacuum energy as predicted by quantum field theory and  the effective cosmological constant determined through astronomical observations. The objection that inflation would not proceed as in the  standard theory in this case has been shown to be unjustified \cite{ellis2}. \\

Our interest lies in probing the question of whether the trace-free Einstein (TFE) gravity can support compact objects. Clearly the field equations of TFE and the general relativity gravity equations (GRG) are equivalent. However, the presentation of the equations and the interplay between the dynamical and geometrical quantities is expressed differently. Therefore it is interesting to ask:  What is the prognosis for the existence of bounded distributions in this scenario? In standard Einstein theory, the field equations for spherically symmetric matter turns out to be three equations in four unknowns: two gravitational potentials and the density and pressure. To close the system an assumption must be made - historically {\it {ad hoc}} forms for one of the geometrical or dynamical variables have been made or an equation
of state relating two of the quantities is imposed. The conservation equations following from Ostrogradsky's theorem \cite{Woo15} convey no new information and may be used in place of any of the field equations. In TFE theory the field equations amount to two field equations in four unknowns. The vanishing divergence of the energy momentum tensor is no longer guaranteed and this
 condition must be imposed by hand as a constraint on the system thus bringing us back to 3 equations in 4 unknowns. This means that one of the variables must be assumed to have a particular form or an equation of state can be proposed. Note that the TFE field equations are equivalent to the standard Einstein's field equations, however, the presentation of the equations is manifestly different. In particular, it is well known that the density and pressure separate in the field equations of the standard theory however it is found that in the TFE theory the density and pressure are inextricably linked and occur as a sum known as the inertial energy density ($\rho + p$).  Moreover, the $G_{00} =T_{00}$ field equation in Einstein theory expresses the density in terms of a single gravitational potential, whereas in the TFE scenario the $\hat{G}_{00}=\hat{T}_{00}$ yields a partial differential equation in the two metric potentials as well as the pressure and density. Nevertheless, it is still interesting to investigate exact solutions in this framework to see if any new information emerges.

    The mathematical complexity of the problem is noteworthy. There exists no theory of partial differential equations and consequently systems of partial differential equations. If we have a system of $q$ coupled nonlinear partial differential equations in $q+1$ unknowns, it is not necessarily true that if any set of $q$ unknowns are determined then the remaining unknown is uniquely determined. That is, the solution space is not unique.  It shall be demonstrated through a number of examples that while the metric potentials of GRG still hold in the TFE, different dynamical profiles emerge.  This may be attributed to the nonlinearity of the field equations.

    The static fluid sphere has a lengthy history starting with the Schwarzschild interior solution of 1916. Since then upwards of a hundred exact solutions have been reported in the literature. For example see the compendium of Stephani {\it {et al}} (\cite{stephani}). A comprehensive list also exists in Finch and Skea \cite{finch}. It was shown by Delgaty and Lake \cite{delgaty} that of the solutions reported only a small subset of fewer that 10 satisfied elementary physical requirements and could be used to model relativistic stars. An important such solution was due to Finch and Skea (\cite{fs}) and it was amply demonstrated that the model so generated satisfied was consistent with observations in astrophysics.

During the past decade several solution generating algorithms for constructing solutions to the field equations were reported. Initially Wyman (\cite{wyman}) proposed such an algorithm in 1949 but the ideas were only revived recently. Essentially all the algorithms rely on being able to perform an integration at some stage - and this is where the algorithm often falters. Nevertheless, the algorithms do represent some advance from the ad hoc approaches attempted historically.  Fodor (\cite{fodor}) described a procedure involving a single integration - however the caveat was that the integrand contained a square root by design. Rahman and Visser (\cite{RV}) then devise a generating function using isotropic coordinates and this process also involves a single integration. Lake (\cite{lake}) revives the ideas of Wyman and produces a recipe for curvature and isotropic coordinates but with the familiar weakness of requiring integrations that are nontrivial. Novel solutions are reported using this algorithm. Martin and Visser (\cite{MV}) propose a procedure for Schwarzschild coordinates.  Boonserm, Visser and Weinfurtner (\cite{BVW}) developed theorems to map exact solutions for perfect fluid spheres to other spheres and this exercise is useful in classifying the over 100 exact solutions reported.  Finally Hansraj and Krupanandan (\cite{hans-krup}) exploited a coordinate transformation that allowed the master isotropy equation to be integrated in general. However, the tangible success of the algorithm rests on the ability to actually perform the required integration based on prescribing a single function.

Exact models of static spherically symmetric fluids with a cosmological constant were studied by Winter (\cite{winter}) who considered the Tolman--Oppenheimer--Volkoff (TOV) equation together with some assumptions on the equation of state.  B\"{o}hmer (\cite{bohmer}) examined the TOV equation with a cosmological constant and reported 11 new solutions for the case of a constant energy density. Note that the TOV equation is obtained by incorporating the gravitational mass as well as the mean density. In our work, we do not have an actual cosmological constant - it disappears on account of using the trace-free equations of Einstein. Nevertheless we show that despite its absence, the cosmological constant does leave a footprint in the solutions. Also note that we utilise the conservation equations in the standard form without moving to the TOV equation.

In this paper we derive the appropriate trace-free Einstein equations supplemented by the conservation equation. We describe a procedure to determine exact solutions of the system of field equations. A few examples are examined and the process is demonstrated. In particular we analyse the Schwarzschild geometrical ansatz, the constant density assumption, the case of an isothermal fluid, postulating a constant gravitational potential and finally we examine the Finch-Skea ansatz \cite{fs} in this context. It is shown that compact distributions may indeed exist in this framework and reasonable basic conditions for physical plausibility are satisfied. This includes the existence of a vanishing pressure hypersurface, satisfaction of the energy conditions and causality.

%\subsection{}
%\subsubsection{}

\section{Trace free Einstein gravity}

 A detailed exposition of trace-free Einstein gravity and its relationship with unimodular gravity \cite{unimod1,unimod2,unimod3} is given in \cite{ellis1} and \cite{ellis2}. Nevertheless, we mention a few salient points. Recall that in the standard formulation the gravitational field is governed by the Einstein field equations
 \be
 G_{ab}:= R_{ab} - \frac{1}{2} R g_{ab} = T_{ab}  \label{EFE}
 \ee
where we have set $G=c=1$ using geometrized units. The conservation equations follow naturally: \begin{equation}\label{cons}
G^{ab}_{\,\, ;b}=0 \Rightarrow  T^{ab}_{\,\, ;b}=0.
\end{equation} Denoting by a hat the trace-free part of a symmetric tensor, we may write
\be
\hat{G}_{ab} = R_{ab} - \frac{1}{4}Rg_{ab} \,,\hspace{0.5cm}  \hat{T}_{ab} = T_{ab} -\frac{1}{4} Tg_{ab} \hspace{0.5cm} \Rightarrow \hspace{0.5cm} \hat{G}^a_a = 0, \,\, \hat{T}^a_a=0
\ee
then (\ref{EFE}) implies
\be \label{TFE}
\hat{G}_{ab} = \hat{T}_{ab} \hspace{0.5cm} \Leftrightarrow \hspace{0.5cm} R_{ab} -\frac{1}{4}Rg_{ab} = T_{ab} - \frac{1}{4} Tg_{ab}
\ee
which are the trace -free Einstein field equations (TFE). These are now the equations of motion we use for the gravitational field. Observe that the conservation laws \begin{equation}\label{cons}
T^{ab}_{\,\, ;b} = 0
\end{equation} no longer follow as a natural consequence of the field equations. They must be inserted to the system as an additional constraint. Then taking the divergence of (\ref{TFE}) and integrating gives
\be
 G_{ab} + \Lambda g_{ab} = T_{ab}  \label{EFE1}
 \ee
where the cosmological constant $\Lambda$ is a constant of integration that has nothing to do with vacuum energy \cite{weinberg,ellis1}. It must have a very small value in order to be compatible with solar system observations and  compact star solutions. Thus TFE solutions are the same as GRG solutions with an arbitrary cosmological constant (which can even be zero).

\section{TFE Field Equations}

The static spherically symmetric spacetime in coordinates $(t, r, \theta, \phi)$ is taken as
\begin{equation}
ds^2 = -e^{2\nu(r)}dt^2  + e^{2\lambda (r)} dr^2 + r^2 \left(d\theta^2 +\sin^2 \theta d\phi^2\right) \label{1}
\end{equation}
where the gravitational potentials $\nu$ and $\lambda$ are functions of the radial coordinate $r$ only. We utilise a comoving fluid 4-velocity $ u^a = e^{-\nu} \delta_a^0 $ and  a perfect fluid source with energy momentum tensor $T_{ab} = (\rho + p)u_a u_b + p g_{ab}$ in geometrized units setting the gravitational constant $G$ and the speed of light $c$ to unity. The quantities $\rho$ and $p$ are the energy density and pressure respectively.

The trace free Einstein tensor reads as
\begin{eqnarray}
\hat{G}_{tt} &=& \frac{e^{2(\nu - \lambda)} }{2r^2} \left(r^2(\nu''+\nu'^2 - \nu'\lambda') +2r(\nu' + \lambda') + e^{2\lambda} - 1 \right)  \label{2a} \\ \nonumber \\
\hat{G}_{rr} &=& \frac{1}{2r^2} \left(2r(\nu' + \lambda') -r^2(\nu'' + \nu'^2 -\nu'\lambda') -e^{2\lambda} + 1\right)  \label{2b} \\ \nonumber \\
\hat{G}_{\theta \theta} &=& \frac{e^{-2\lambda}}{2}\left(r^2(\nu'' + \nu'^2 - \lambda'\nu') + e^{2\lambda} -1 \right)  \label{2c} \\ \nonumber \\
\hat{G}_{\phi \phi} &=& \sin^2 \theta G_{\theta \theta}. \label{2d}
\end{eqnarray}
%and in contrast the standard Einstein tensor has the form
%\begin{eqnarray}
%G_{tt} &=& \frac{e^{2(\nu-\lambda)}}{r^2}\left(2r\lambda' + e^{2\lambda} -1\right) \\ \nonumber \\
%G_{rr} &=& -\frac{1}{r^2} \left( e^{2\lambda} -1 -2r\nu'\right) \\ \nonumber \\
%G_{\theta \theta} &=& -re^{-2\lambda} \left(\lambda' - \nu' -r\nu'' +r\lambda'\nu' -r\nu'^2\right) \\ \nonumber \\
%G_{\phi \phi} &=& \sin^2 \theta G_{\theta \theta}
%\end{eqnarray}

The trace-free components of the energy-momentum tensor are
\begin{equation}
\hat{T}_{ab} = \left( \frac{3}{4}(\rho + p)e^{2\nu}, \frac{1}{4}(\rho + p)e^{2\lambda}, \frac{r^2}{4} (\rho + p), \frac{r^2 \sin^2 \theta}{4} (\rho + p)\right) \label{3}
\end{equation}
from which the coupling of density ($\rho$) and pressure ($p$) is readily apparent. Ordinarily in the regular Einstein field equations the $G_{00}$ and $T_{00}$ components are free of pressure.

The TFE field equations now have the form
\beq
  (\nu'' + \nu'^2   - \nu'\lambda' ) +\frac{2}{r}\left(\nu' + \lambda'\right) +\frac{e^{\lambda}-1}{r^2}     &=&\frac{3}{2}(\rho + p)e^{2\lambda} \label{4a} \\ \nonumber \\
 \frac{2}{r}\left(\nu' + \lambda'\right)  -(\nu''  +\nu'^2 - \nu' \lambda') -\frac{e^{2\lambda} -1}{r^2} &=& \frac{1}{2}(\rho + p)e^{2\lambda}  \label{4b} \\ \nonumber \\
 (\nu'' + \nu'^2 - \nu'\lambda') + \frac{e^{2\lambda}-1}{r^2}&=& \frac{1}{2}(\rho + p)e^{2\lambda} \label{4c}
\eeq
It is easy to verify that these three equations simultaneously imply the master equation
\be
r^2(\nu'' + \nu'^2 - \nu'\lambda') -r(\nu'+\lambda') +(e^{2\lambda}-1) = 0 \label{4d}
\ee
which is also taken as the equation of pressure isotropy, and the further equation
\begin{equation}
\frac{2}{r}\left(\nu' + \lambda'\right) = (\rho + p)e^{2\lambda}.
\label{4d1}
\end{equation}
The two equations (\ref{4d}),(\ref{4d1}) imply all three of (\ref{4a})-(\ref{4c}). Finally the conservation equations (\ref{cons})
 reduce to the equation
\be
p' + (\rho + p)\nu' = 0
                \label{4e}
\ee
which  is to be added as an additional assumption to the system, as discussed above  (\cite{weinberg,ellis1}).
Thus the TFE equations are satisfied if we satisfy (\ref{4d}), (\ref{4d1}), (\ref{4e}).
If a metric is found that satisfies (\ref{4d}) then the quantity $\rho + p$ may be obtained by backward substitution from (\ref{4d1}).
Note that  it would be impossible to find the density and pressure independently from (\ref{4d}) and (\ref{4d1}).
Later, we show how this may be achieved via (\ref{4e}).
%One can put the same
%$p = p(\rho)$ in (\ref{4e}). See below}]\\

Birkhoff's theorem guarantees that the exterior field of any static or non--static spherical distribution is the Schwarzschild exterior metric. Obviously this holds here in view of the spherical geometry.  We easily deduce this from the field equations. From $\hat{G}_{tt}=0$ and $\hat{G}_{rr} =0$ we immediately get that $\lambda' + \nu'= 0$. Using this result in $\hat{G}_{\theta \theta} =0$ gives the familiar $e^{2\lambda} = \left(1-\frac{2M}{r}\right)^{-1}$ where $M$ is a constant associated with the gravitational mass of the perfect fluid ball. The corresponding temporal potential then has the usual form $e^{2\nu} = 1-\frac{2M}{r}$. It should also be remembered that  the normal Israel--Darmois junction conditions still hold in trace-free Einstein theory  \cite{ellis1}. That is pressure vanishes for a radial value and the components of the metric tensor of the interior and exterior spacetime match.\\

 The transformations $x=Cr^2$ ($C$ a constant),  $e^{2\nu(r)} = y^2(x)$ and $e^{-2\lambda (r)} = Z(x)$ are known to convert the master isotropy equation in Einstein gravity to a linear second order differential equation. Applying these transformations to (\ref{4d}) renders it in the form
 \be
 4x^2Z\ddot{y} + 2x^2\dot{Z}\dot{y} + (\dot{Z}x - Z +1) y = 0 \label{4f}
 \ee
where the dots denote derivatives with respect to $x$. This is exactly the same as the  equation of pressure isotropy in standard Einstein gravity \cite{chil-hans}.
 With the aforesaid transformations, (\ref{4d1}) becomes
 \begin{equation}
4Z\dot{y}-2\dot{Z}y=\frac{\rho+p}{C}y \label{4f1}
 \end{equation}
 and (\ref{4e}) assumes the form
\be
\dot{p} + (\rho + p)\frac{\dot{y}}{y} =0 \label{4e2}
\ee
which is the energy conservation equation applicable here.

\section{Linear Equation of State}

As the system is under-determined, the most physically important way to close the system of field equations is to assume an equation of state. This approach has not necessarily led to exact solutions in the standard GRG for static spheres. For example, Nillson and Uggla \cite{Nils1,Nils2} have considered both a linear and polytropic equation of state and the treatment was completed using numerical methods. They also employed a dynamical systems approach and other useful information was extracted, but not an exact solution. Might we have some success in attempting an equation of state in the context of the TFEs? The answer turns out to be in the negative as shall be shown below.  In our case, assuming a barotropic equation of state $p = p(\rho) $ results in the relationship
\be
(\rho(p)+p)\dot{ }= \left(1+ \frac{d\rho}{dp}\right)\dot{p} = -  (\rho(p) + p)\frac{\dot{y}}{y} \label{4m1}
\ee
and specialising to the linear equation of state $p = w \rho$ where $w \neq -1$ is constant, then (\ref{4m1}) yields
\be
\frac{(\rho+p)\dot{}}{\rho+p} =   -\frac{w+1}{w}\frac{\dot{y}}{y} \Rightarrow
\log (\rho+p) = -\frac{w+1}{w} \log y +C \label{4m1a}
%\rho(r) = \frac{C}{y(r)}
\ee
so
\be
\rho = c_1 y^{-\frac{w+1}{w}} \label{4m2}
%\rho(r) = \frac{C}{y(r)}
\ee
and equation (\ref{4f1}) assumes the form
 \begin{equation}
4Z\dot{y}-2\dot{Z}y=\frac{c_1(w+1)}{C}y^{-\frac{1}{w}} \label{4m7}
 \end{equation}
on substituting (\ref{4m2}). Invoking (\ref{4m2}) then differentiating (\ref{4f1}) with respect to $x$ and multiplying throughout by $x^2$, we obtain
\be
4x^2Z\ddot{y} + 2x^2\dot{Z}\dot{y} - 2x^2\ddot{Z}y=-\frac{(w+1)c_1}{wC} x^2 y^{-\frac{1}{w}-1} \dot{y} \label{4m3}
\ee
Subtracting (\ref{4m3}) from (\ref{4f}) results in
\be
\frac{wC}{c_1 (w+1)x^2}\left(\dot{Z}x -Z+1+2x^2\ddot{Z}\right) = y^{-\frac{1}{w}-2}\dot{y} \label{4m4}
\ee
and we have succeeded in separating the gravitational potentials $Z$ and $y$. Now plugging $\dot{y}$ from (\ref{4m4}) into (\ref{4f1}) gives
\be
4Z\left(\frac{wC}{c_1(w+1)}\right) \left(\frac{\dot{Z}x-Z+1+2x^2\ddot{Z}}{x^2}\right)y^{\frac{2}{w}+2} -2\dot{Z}y^{\frac{1}{w}+1} -\frac{(w+1)c_1}{C}=0 \label{4m5}
\ee
after using (\ref{4m2}). Note that (\ref{4m5}) is quadratic in $y$ so we may write
\be
y= \left(-\frac{c_1 (w+1) x \left(x \dot{Z} \pm \sqrt{4 w Z \left(2 x^2 \ddot{Z}+x \dot{Z}+1\right)-4 w Z^2+x^2 \dot{Z}^2}\right)}{4C w Z \left(-2 x^2 \ddot{Z}-x \dot{Z}+Z-1\right)}\right)^{\frac{w}{w+1}} \label{4m6}
\ee
which is now explicitly in terms of $Z$ and its derivatives.
Substituting in (\ref{4m7}) generates the differential equation
\beq
\frac{C w h}{c_1 (w+1) x^2}  - \left(w \left((x\dot{Z}-f)\left(Zh-x\dot{Z}h -2x^3Z^4 -5x^2Z\dot{Z}  \right) \right. \right. && \n \\ \n \\
+  \left. \left.  \frac{x Z h}{f} \left(\left(\dot{Z} \left(-(4 w+1) x^2 \ddot{Z}+2 w (Z-1)\right)-x(2 w +1) \dot{Z}^2-2 w x Z \left(2 x Z^{3}+5 \ddot{Z}\right)\right)   \right. \right. \right. && \n \\ \n \\
\left. \left. \left.+x \ddot{Z}+\dot{Z}\right)\right)
  \frac{4CwZh}{c_1(w+1)x(x\dot{Z}-f)}\right) / \left((w+1) x Z \left(h\right) \left(x \dot{Z}-f\right)\right) &=& 0 \n \\ \label{4m7}
\eeq
where $f= \sqrt{4 w Z \left(2 x^2 \ddot{Z}+x \dot{Z}+1\right)-4 w Z^2+x^2 \dot{Z}^2}$ and $h= 2 x^2 \ddot{Z}+x \dot{Z}-Z+1 $. A solution to (\ref{4m7}) will generate an exact model of the TFEs for a static spherically symmetric distribution of perfect fluid with a linear equation of state. However, we are unable to find such a solution on account of the intractable differential equation (\ref{4m7}). This illustrates the difficulty of finding exact solutions by imposing physical conditions {\it{ a priori}}.

\section{Solution generating algorithm}

The procedure in solving the system is as follows: First we choose one of $Z$ or $y$ in (\ref{4f}) and by integration find the other. These two are then plugged into  (\ref{4d1}) to give the inertial mass density $\rho + p$. This last quantity is then substituted into (\ref{4e2}) to explicitly reveal $p(r)$ provided the integration is tractable. Thereafter the density $\rho(r)$ may be found by subtracting the pressure $p$ from $\rho + p$. Now the model is completely determined.

  We are at liberty to specify one of $Z$ or $y$ {\it a priori} and by integration find the other potential. This approach has been exhausted for the Einstein gravity case and in recent times various equivalent algorithms have been developed  \cite{lake, RV, MV, BVW, hans-krup}. Indeed, since the pressure isotropy is preserved, any metric representing a perfect fluid source may be used.  All solutions found this way may then be plugged back into any one of (\ref{4f1}) or (\ref{4e2})  to generate  $\rho + p$. %For this purpose note that (\ref{4c}) may be written as
 %\ %be
 %\rho + p = %\frac{2C}{xy}\left[4x^2Z\ddot{y} + 2x(Z+x\dot{Z})\dot{y} + (1-Z)y \right] \label{4e3}
 %\ee
 %and with the help of (\ref{4f}), this is further simplified to
 %\be
 %\rho + p = \frac{2C}{y}\left(2Z\dot{y} - \dot{Z}y\right)  \label{4e4}
 %\ee
 %or
 %\be
 %\rho + p = \frac{2e^{-2\lambda}}{r}\left(\nu' + \lambda'\right) \label{4e5}
 %\ee
 %in the canonical variables.
  While any perfect fluid metric may be utilised, the caveat in this programme is that the calculation of the pressure may be thwarted by our inability to integrate (\ref{4e2}).

 % [\textcolor{red}{Alternative?? Choose $p=p(\rho)$ and integrate ?? See above}]

Let us illustrate this procedure  with some examples.

\section{Incompressible fluid sphere}

Since the earliest compact star model emerged from the assumption  of a constant energy density to yield the Schwarzschild interior metric, it is interesting to examine its antitype in the trace-free scenario.   Setting $\rho = \rho_0$ a constant, equation (\ref{4e2}) immediately yields
\be
\rho_0 + p = c_1 y^{-1} \label{4k1}
\ee
($c_1$ is an integration constant) expressing the pressure in terms of one of the gravitational potentials. At this point we should point out that this choice of the energy density is fortuitous to allow for the integration of the field equations. It turns out that prescribing the density {\it a priori} is not a viable option in general since the density and pressure are coupled as the inertial mass density $\rho + p$ in this scenario. In the ordinary Einstein gravity case the $g_{tt}$ components contain only the density $\rho$ and potential $\lambda$ thus prescribing any one is tantamount to selecting the other. We do not have that luxury with the trace-free equations of gravity.  Besides the constant density case, there appears no functional expressions for the density or pressure that readily allow for the complete revelation of the entire model. It appears unavoidable, at least using the coordinate transformation we have employed, to prescribe the geometry $\nu$ and $\lambda$ at the outset. Specifying the inertial mass density $\rho + p$ may also work, however, the physical meaning in general is not clear. For specific case such as the isothermal sphere $\rho$ and $p$ both behaving as $\frac{1}{r^2}$ this possibility is examined later.

Using (\ref{4k1}   equation (\ref{4d1}) becomes
\be
\rho_0 + p = \frac{2C}{y}\left(2Z\dot{y} - \dot{Z}y\right) \label{4k2}
\ee
and together with (\ref{4k1}) gives the constraint
\be
2Z\dot{y} - \dot{Z}y=\frac{c_1}{2C} \label{4k3}
\ee
Differentiating (\ref{4k3}) with respect to $x$ and multiplying throughout by $2x^2$ gives
\[
4x^2 Z\ddot{y}+2x^2\dot{Z}\dot{y} -2x^2 \ddot{Z}y = 0
\]
Comparing with the pressure isotropy condition (\ref{4f}) gives a differential equation only in $Z$ written as
\be
2x^2\ddot{Z} + \dot{Z}x -Z+1=0 \label{4k4}
\ee
which is solved by
\be
Z=1+c_2 x + \frac{c_3}{\sqrt{x}} \label{4k5}
\ee
Note that this is not the same as the Schwarzschild potential  in general, it takes that form only if $c_3 = 0$, since $c_3$ represents the cosmological constant as follows from the discussion above, see (\ref{EFE1}). Thus this potential is of the de Sitter--Schwarzschild type and is related to the  Nariai \cite{nariai} solution  which occurs in the Einstein gravity with a non-zero cosmological constant. The remaining potential is established via (\ref{4k3}) and has the form
\be
y= \sqrt{1+c_2 x + \frac{c_3}{\sqrt{x}}}\left(c_4 + \frac{c_1}{4C}\int \left(1+c_2 x + \frac{c_3}{\sqrt{x}}\right)^{-3/2} dx\right) \label{4k6}
\ee
where $c_4$ is a further integration constant. Note that usually in the standard Einstein equations, there exists only two necessary constants of integrations and these are usually settled by invoking the boundary conditions. The presence of extra constants of integration in the present context arises out of the introduction of the conservation law as an extra condition as well as the process of integration.
The potential (\ref{4k6})  may be obtained in terms of elliptic functions which are extremely lengthy.  It is instructive to examine the consequences of one of the integration constants vanishing. If we set $c_3 = 0$, then we  immediately recover the Schwarzschild interior solution. On the other hand setting $c_2 = 0$ yields a completely new geometry. We obtain for the potential
\beq
y &=& \frac{1}{16C \sqrt{x}} \left(c_1 \left(2 \sqrt{c_3 +\sqrt{x}} \left(-15 c_3^2-5 c_3 \sqrt{x}+2 x\right) \right. \right. \nonumber  \\
 && \left. \left. +15 c_3^2 \left(c_3+\sqrt{x}\right) \log \left(2  \left(\sqrt{c_3 + \sqrt{x}}+\sqrt{x}\right)+c_3\right)\right)
+ 16C c_4\sqrt{c_3 +\sqrt{x}}\right) \label{4k7}
\eeq
The pressure evaluates to
\be
p= \frac{16 C \sqrt{x}}{15  c_3^2 \left(\sqrt{c_3\sqrt{x}+x} \log \left(2  \left(\sqrt{c_3\sqrt{x} +x} +1\right)+c_3\right)-2 \right)-10  c_3 x +4  x^{\frac{3}{2}} +16 \frac{c_4 C}{c_1} \sqrt{c_3\sqrt{x}+x}}- \rho_0 \label{4k8}
\ee
While a pressure-free hypersurface may exist demarcating the boundary of this fluid, we neglect to study it further in view of the fact that the assumption of constant density renders the causality criterion invalid.

 %Thus we saw that prescribing the form of the matter in terms of constant density necessarily yields a fluid with unphysical negative pressure in trace free gravity, which is not the case in GR. To see it more transparently, in %the next section we will force the interior metric to be of the same form as the Schwarzschild interior geometry, and see what kind of matter field would have such a solution \textcolor{red}{This cannot be true for $c_3 = 0$ as %this is the GR result. Also the negative pressure might perhaps be avoided if we take into account a finite radius for the surface of the star - indeed $p=0$ might locate that surface for us.}

\section{Interior Schwarzschild geometry}

The choice $Z=1+x$ yields the constant density (incompressible fluid) solution in the Einstein gravity \cite{chil-hans}. We use this postulate to obtain the density and pressure in the current gravity theory. By straightforward integration, the equation (\ref{4d}) yields the remaining potential as
\be
y=c_1\sqrt{1+x} + c_2 \label{4g}
\ee
Note that we have only obtained the Schwarzschild geometry - this does not necessarily mean we have an incompressible sphere as in the Einstein gravity case.
With the aid of (\ref{4c}) we obtain
\be
\rho + p = \frac{- 2Cc_2}{c_1\sqrt{1+x} +c_2} \label{4h}
\ee
When this is substituted into (\ref{4e2}), the pressure integrates out as
\be
p=\frac{-2Cc_2}{c_1\sqrt{1+x}+c_2} \label{4j} - \kappa
\ee
where $\kappa$ is a constant.
Consequently the density may be obtained via (\ref{4h}) and has the value
\be
\rho = \kappa \label{4k}
\ee
which is a constant as expected. That is the interior Schwarzschild geometry does indeed generate an incompressible  fluid and is consistent with the standard Einstein's equations. As shown in the preceding section the converse is not true: the assumption of constant density yields a richer geometry of which the Schwarzschild metric is a special case.

\section{Isothermal fluid}

Isothermal fluid is characterised by the behaviour $\rho \sim \frac{1}{r^2}$ and the linear barotropic equation of state $p = \alpha \rho$ ($0 < \alpha < 1$). Imposing two constraints on a static model in Einstein gravity, which comprises three equations in four unknowns for spherical symmetry, appears to be over-determining the system. This is true, however, the pressure isotropy may then be interpreted as a consistency condition and is used to ascertain unknown constants. This is precisely the path taken by Saslaw {\it et al} \cite{saslaw}. In our case, we have one master equation in two unknowns and a conservation equation that must be satisfied. Accordingly, a different model should emerge.

 In the present discussion we set $\rho = \frac{A}{r^2}$ for some constant $A$ so  $\rho + p = \frac{A(1+\alpha)}{r^2}$. Substituting these in (\ref{4e}) gives
 \be
 e^{2\nu} = (Kr)^{\frac{4\alpha}{1+\alpha}} \label{501}
 \ee
 where $K$ is an inconsequential constant of integration. Invoking (\ref{4d1}) we obtain the expression
 \be
 (1+\alpha)A e^{2\lambda} = \frac{4\alpha}{(1+\alpha)r} + 2\lambda'  \label{502}
 \ee
 With the transformation $e^{2\lambda} = F(r)$ for some function $F$, equation (\ref{502}) assumes the form
 \be
 (1+\alpha)A F^2 = \frac{4\alpha}{(1+\alpha)} F + rF' \label{503}
 \ee
 which is a Ricatti type equation that is readily solvable as
 \be
 F(r) = e^{2\lambda} = A^{-1}\left(\left(C_1 (1+\alpha)r\right)^{\frac{4\alpha}{1+\alpha}} - (\alpha + 1)^2 \right)^{-1} \label{504}
 \ee
for some constant of integration $C_1$.   In contrast the isothermal fluid sphere in Einstein gravity \cite{saslaw} has an identical form for $\nu$ but for a constant value for $\lambda$. In the more general Lovelock gravity theory for all orders of spacetime dimension $d$  and any order of the Lovelock polynomial $N$ it has been shown that $\lambda = $ constant is a necessary and sufficient condition for isothermal behaviour and that the Saslaw {\it {et al}} \cite{saslaw} metric in 4 dimensions is universal for all $d$ and $N$ \cite{NKD}. Clearly that is not the case here in the trace-free scenario. This underscores the fact that the solution of Saslaw {\it {et al}} is not the most general solution and is regained in our model by setting $C_1 =0$. Note by design the isothermal sphere has density and pressure going as $\frac{1}{r^2}$ so  a singularity at the stellar centre exists. Moreover, such distributions do not admit a hypersurface of vanishing pressure so cannot represent compact objects. They may however feature as a shell of isothermal fluid in a multiple layer model such as a gravastar \cite{mazur}.

The preceding case motivates the consideration of the solution for $\lambda $ being a constant since this is known to be a necessary and sufficient condition for isothermal behaviour in Einstein gravity. Let us set $e^{2\lambda}  = 1 + \gamma$ for some constant $\gamma$. Then equation (\ref{4d}) reduces to
  \be
  r^2 \nu'' + r^2\nu'^2 -r\nu' + \gamma = 0 \label{601}
    \ee
and is solved by
\be
e^{2\nu} = c_2 \left(r^{2\sqrt{1-\gamma}} + c_1\right)^2 \left(r^{1-\sqrt{1-\gamma}}\right)^2  \label{602}
\ee
where $c_1$ and $c_2$ are integration constants that may be fixed by the exterior Schwarzschild solution provided a pressure free boundary exists.  Equation (\ref{4d1}) reduces to $\rho + p = \frac{2}{\gamma - 1} \frac{\nu'}{r}$ and when substituted into the conservation equation (\ref{4e}) yields
\be
p= \frac{2 \left(\sqrt{1-\gamma }+1\right)-\gamma -\frac{4 \sqrt{1-\gamma } c_1}{c_1+r^{2 \sqrt{1-\gamma }}}}{(\gamma -1) r^2} \label{603}
\ee
and the density has the form
\be
\rho = \frac{\gamma }{(\gamma -1) r^2} \label{604}
\ee
which is of the isothermal form as in the Newtonian theory however the pressure function is not proportional to the density. Moreover, unlike the GRG case, the pressure function admits a hypersurface of zero pressure establishing the boundary of the compact object. This contradicts the model of Saslaw {\it {et al}} which was valid for an unbounded fluid.  The form of (\ref{604}) suggest that the case $\gamma = 1$ must be treated separately. For this case, the gravitational potential has the form
\be
e^{2\nu} = c_5 r^2 \left(\ln r - c_6\right)^2 \label{605}
\ee
where $c_5$ and $c_6$ are constants. The pressure profile obeys
\be
p= \frac{1}{2r^2} \left(\frac{2}{\ln r- c_5}+1\right)
\ee
while the density has the form
\be
\rho = \frac{1}{2r^2} \label{606}
\ee
which is again of the isothermal form. Interestingly for all $\gamma$ a vanishing pressure hypersurface exists thus demonstrating that a constant gravitational potential generates a compact static configuration of perfect fluid matter.

\section{Finch--Skea ansatz $\displaystyle{Z=\frac{1}{1+x}}$}

 We now consider  the class of metrics given by Finch--Skea \cite{fs} that describes a large class of perfect fluid spacetimes in general relativity. The Finch--Skea \cite{fs} metric was  based originally on an ansatz proposed by Duorah and Ray \cite{Duorah}. It has been demonstrated in \cite{fs} that the model conforms to the astrophysical model of Walecka \cite{wal} and corresponds to a special case of the Vaidya--Tikekar spacetime \cite{VT} when the spheroidal parameter vanishes.  With the simpler Finch--Skea choice $Z=\frac{1}{1+x}$
equation (\ref{4f}) is solved by
\be
y=c_1\left(v\cos v - \sin v\right) + c_2\left(v\sin v +\cos v \right)  \label{4l}
\ee
where we have made the transformation $v=\sqrt{1+x}$ following \cite{fs}.
Then we obtain
\be
\rho + p = \frac{2 C \left(\tan v \left(c_2 v^2 -c_1 v (v^2 +1)\right)+ \left(c_1 v^2 + c_2 (v^2 + 1) v\right)\right)}{v^5 \left(\tan v \left(c_2 v- c_1\right)+ \left(c_1 v +c_2\right)\right)} \label{4m}
\ee
Remarkably, the pressure may be obtained explicitly as
\be
\frac{p}{C} =- \frac{1}{v^2} \frac{ \left(\tan v \left((\beta^2 +1) v + \beta  (v^2 + 1)\right)+   \left(\beta^2 v^2 -1\right)\right)}{\left(\beta v + 1\right) \left(\tan v \left( v -\beta\right)+ \left(\beta v + 1\right)\right)} \label{4n}
\ee
where we have set $\beta = \frac{c_1}{c_2}$ following \cite{fs}. This may be compared to the Finch-- Skea pressure of Einstein gravity given by
 \be
 \frac{p}{C} =-\frac{1}{v^2} \frac{(\beta v + 1) +(\beta -v) \tan v}{(\beta v - 1) - (\beta + v)\tan v}
 \ee
 The density has the simple form
\be
\rho=\frac{C (v^2 + 2)}{v^4}  \label{4p}
\ee
and is remarkably identical to the expression obtain in \cite{fs} in the Einstein gravity. Note however, that the pressure profile is different. That is the metric potentials and density match but not the pressure. Using the formula \be
m =\int \rho r^2 dr = \frac{1}{2C^{3/2}} \int \rho \sqrt{x} dx
\ee
we obtain the mass of the star as
\be
m= \frac{1}{8C^{3/2}} \left(\frac{8 x^{3/2}}{x^2+1}-3 \sqrt{2} \log \frac{\left(x+\sqrt{2x}+1\right)}{\left(x-\sqrt{2x}+1\right)} + 6 \sqrt{2} \tan ^{-1}\frac{\sqrt{2x}}{1-x} \right)
\ee
From (\ref{4p}) we may solve for $v$ and substitute in (\ref{4n}) thus obtaining a barotropic equation of state $p = p(\rho)$ which is consistent with distributions of perfect fluid.

The sound speed index is given by
\beq
\frac{dp}{d\rho} &=& \left( v \left( 1-\beta \tan v \right) \left(\tan v \left(-\beta^3 (v^2 +1) v^3 -\beta^2 (v^2 + 2) v^2 +\beta (v^2 -1 ) v + v^2\right)  \right. \right. \nonumber \\ && \left. \left.
+ \left(\beta^3 v^4 + \beta^2 (v^2 + 3)v^3 +\beta (2 v^2+3) v^2+ (v^2 + 1) v\right)\right)\right) \nonumber \\ &&
/ \left((v^2 +4) \left(\beta v + 1\right)^2 \left(\tan v \left( v -\beta\right)+ \left(\beta v +1\right)\right)^2 \right) \label{4q}
\eeq

The expressions above do not lend themselves to an analytic treatment therefore we use graphical plots to investigate the profiles of the dynamical quantities. There are in fact two free parameters $\beta$ and $C$ that must be chosen at the outset to allow to generate plots. In order to constrain these two parameters let us examine the behaviour at the centre $r=0$. We obtain
\beq
\rho_0 &=& 3C   \label{4ra} \\ \nonumber \\
p_0 &=& \frac{- C\left(\tan 1 (\beta +1)^2 +(\beta^2 - 1)\right)}{(\beta + 1)(\tan 1 (1-\beta)+(1+\beta))} \label{4rb} \\ \nonumber \\
\left(\frac{dp}{d\rho}\right)_0 &=& \frac{(1-\beta \tan 1)(\tan 1 (1-3\beta^2 -2\beta^3) +(2+\beta +4\beta^2 + \beta^3))}{5(\beta+1)^2(\tan 1 (1-\beta) + (1+\beta))^2} \label{4rc}
\eeq
where the subscript $_0$ refers to the centre. The condition $\rho_0 > 0$ immediately requires $C>0$. Now a positive central pressure demands that
\be
-1 < \beta < \frac{1-\tan 1}{1+\tan 1} \hspace{0.5cm} {\mbox {or}} \hspace{0.5cm} \beta > \frac{1+\tan 1}{-1+ \tan 1} \label{4s}
\ee
Requiring $\left(\frac{dp}{d\rho}\right)_0 > 0$ gives
\be
\beta < \cot 1 \hspace{0.5cm}  {\mbox {or}} \hspace{0.5cm} \frac{2+\tan 1}{-1 + 2\tan 1} < \beta < \infty \hspace{0.5cm} {\mbox {but}} \hspace{0.5cm} \beta \neq \frac{1+ \tan 1}{-1 + \tan 1} \label{4t}
\ee
while $\left(\frac{dp}{d\rho}\right)_0 < 1$ is satisfied by
\be
\frac{-\sqrt{21}-2 \sin 2-9 \cos 2}{8-9 \sin 2+2 \cos 2} < \beta < \frac{\sqrt{21}-2 \sin 2-9 \cos 2}{8-9 \sin 2+2 \cos 2} \label{4u}
\ee
The inequalities (\ref{4s}), (\ref{4t}) and (\ref{4u}) are simultaneously satisfied in the window
\be
-1 < \beta < \frac{1-\tan 1}{1+\tan 1} \label{4v}
\ee
which when expressed as a decimal is $-1 < \beta < -0,218$.  Accordingly we make the choice $\beta = -0.5$. (This may be contrasted with the Finch-Skea case in Einstein gravity which demanded $0.218 < \beta < 6.407$ as the interval of validity for the central physical conditions to be satisfied. For the charged Finch--Skea analogue the value $\beta = 1$ was chosen in the work of \cite{hans-mah}).  The plots appearing hereunder have been constructed using the parameter values
$C=1.5$ and $\beta = \frac{c_1}{c_2} = - 0.5$.

\begin{figure}
  \includegraphics[width=7cm]{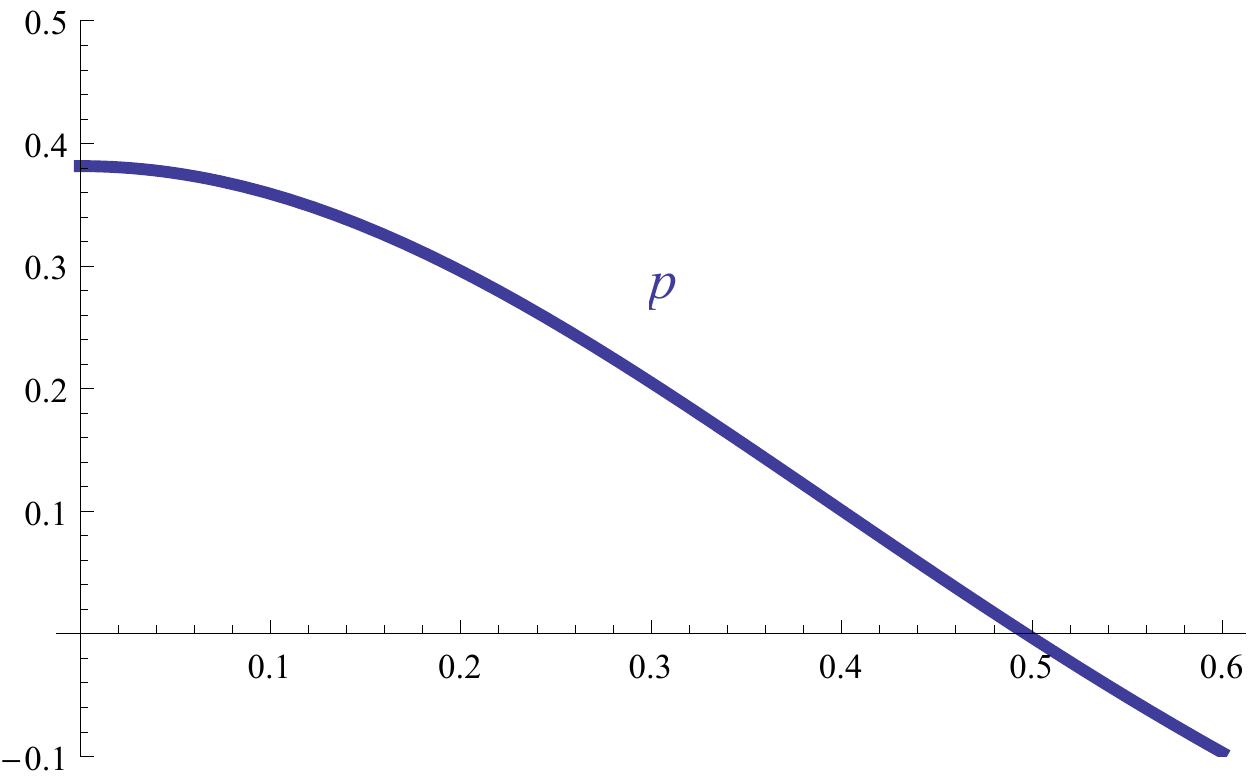}\\
  \caption{Plot of pressure versus radius ($x$)}\label{5}
\end{figure}

\begin{figure}
  \includegraphics[width=7cm]{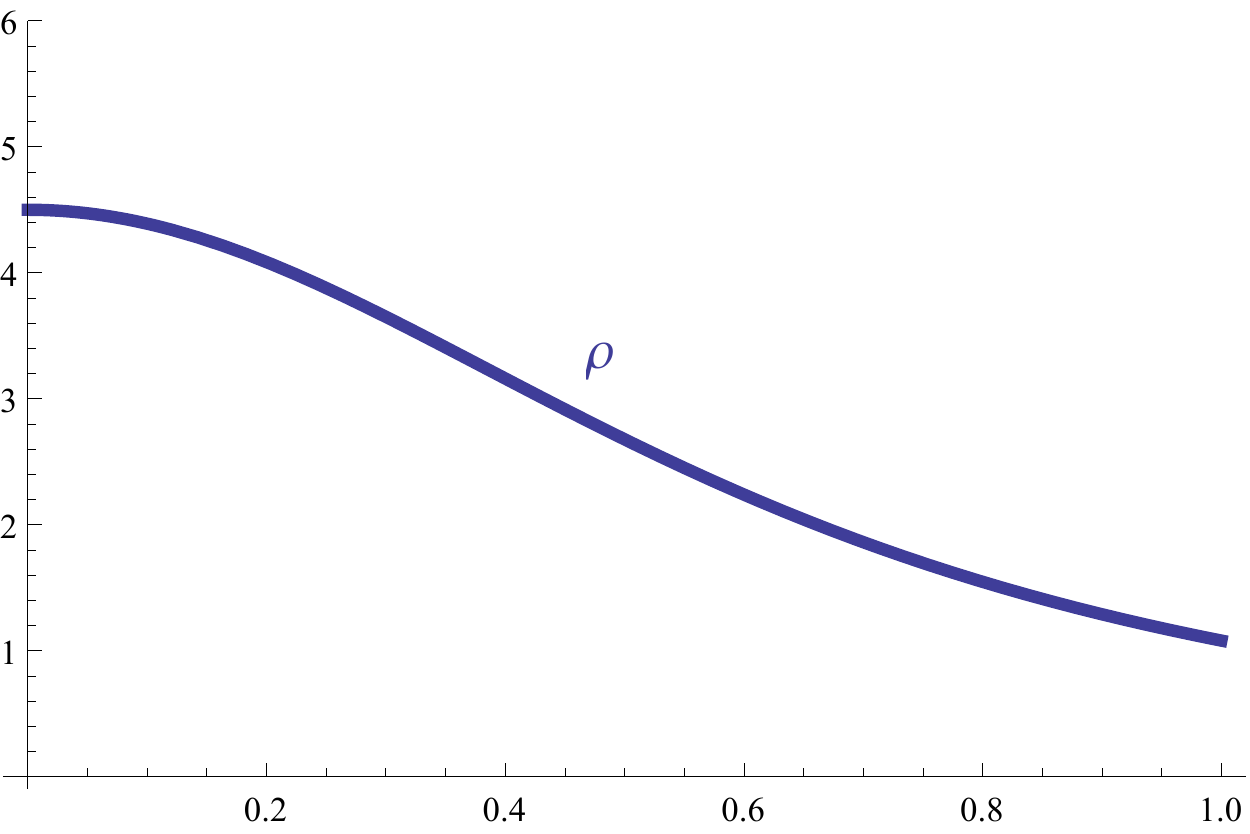}\\
  \caption{Plot of energy density versus radius ($x$)}\label{6}
\end{figure}

\begin{figure}
  \includegraphics[width=7cm]{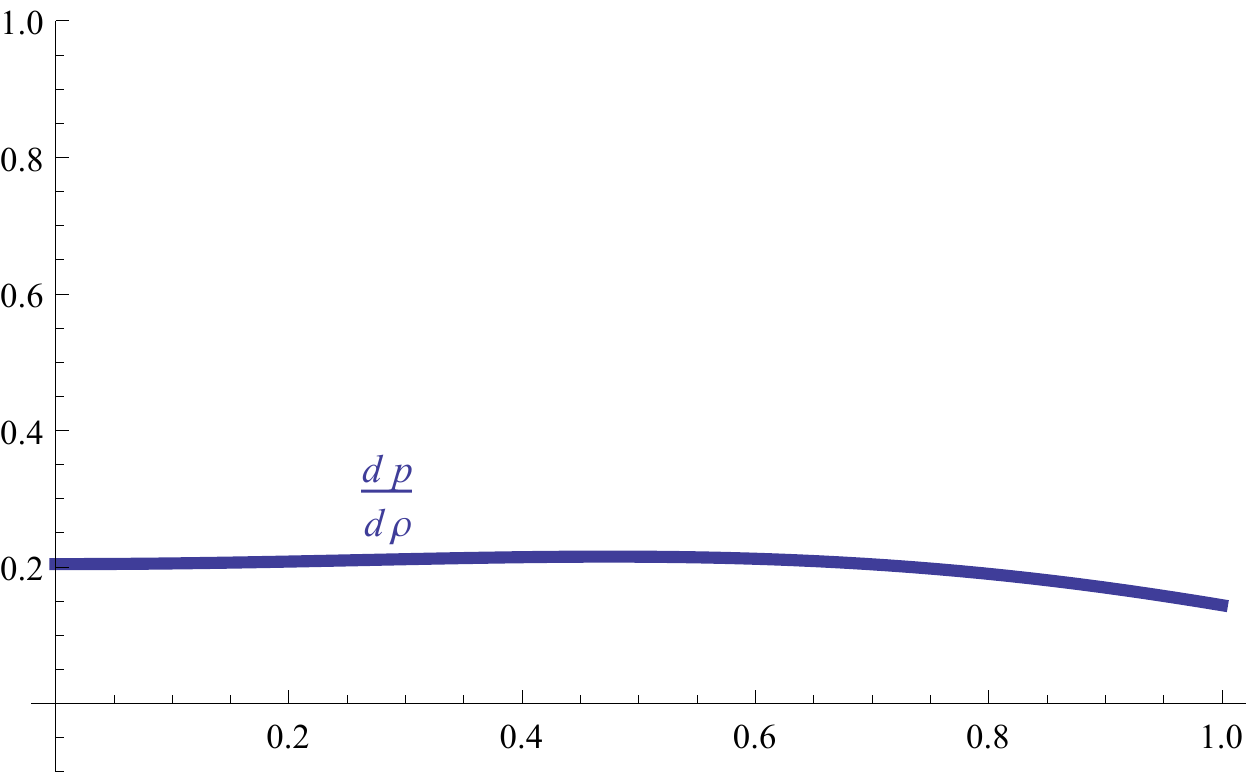}\\
  \caption{Plot of sound-speed index versus radius ($x$)}\label{7}
\end{figure}

\begin{figure}
  \includegraphics[width=7cm]{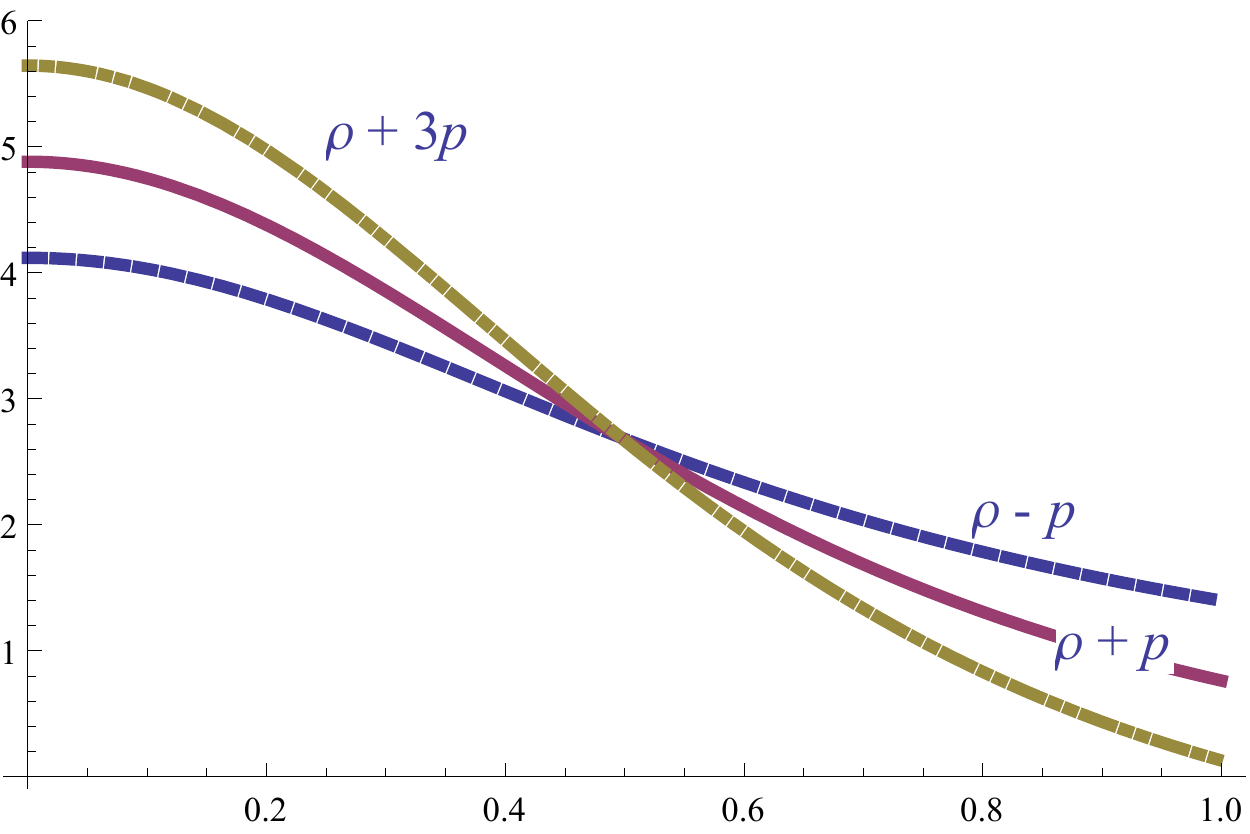}\\
  \caption{Plot of energy conditions versus radius ($x$)}\label{8}
\end{figure}

\begin{figure}
 \includegraphics[width=7cm]{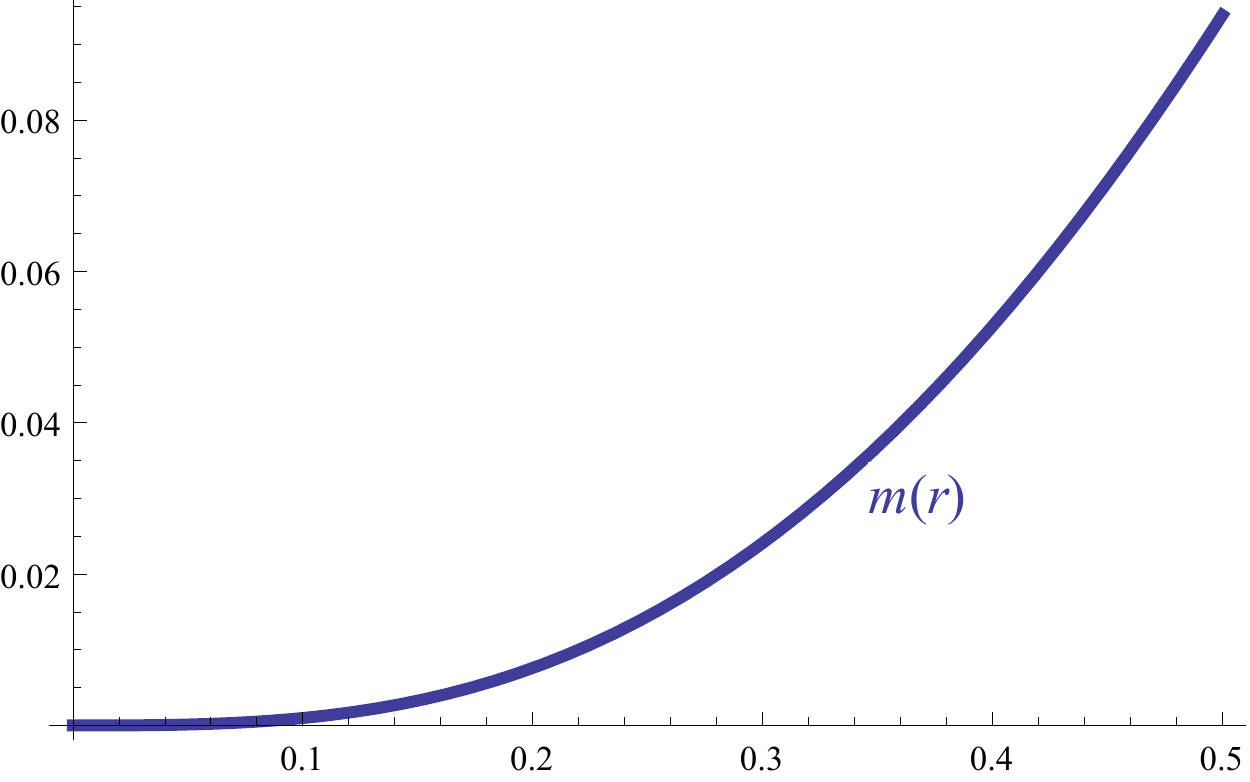}\\
\caption{Plot of mass function versus radius ($x$)}\label{9}
\end{figure}

The plots of the dynamical quantities display a number of pleasing features consistent with realistic distributions of perfect fluid matter. Fig 1 depicts the pressure which is everywhere positive and vanishes for a radial value of $x=0.5$ and is also monotonically decreasing from the centre. Note that the pressure function is periodic so it is necessary to take the first zero of the numerator as the boundary value.   The density (Fig. 2) similarly is positive and decreasing. The speed of sound (Fig. 3) is everywhere less than the speed of light so that causality is not violated. The energy conditions (Fig. 4) are all satisfied for this choice of parameter values. From the boundary value $x = 0.5$ units the radius $R$ may then be established via $x=CR^2$ as 0,577 units. The plot of the mass function (Fig. 5) is smooth and increasing as expected. Moreover, it can be observed that the Buchdahl \cite{Buch} limit $\frac{M}{R} < \frac{4}{9}$ is also satisfied for any radial value.

It now remains to complete the matching across the common hypersurface between the interior and exterior spacetimes. Because of the quadratic terms $\beta^2 = (c_1/C_2)^2$ in the numerator of the pressure there are two possible sets of integration constants. Vanishing of the pressure and the continuity of $g_{rr}$ across the boundary interface $r=R$ gives
\be
\left\{c_1 = -\frac{\sqrt{\frac{R-2 M}{R}} }{(V^2+1)\sin V}, c_2 = -V c_1\right\},\left\{c_1 = -\frac{\sqrt{\frac{R-2 M}{R}} (V \tan V-1)}{2 V \sec V}, c_2 = \frac{V+\tan V}{1-V\tan V}c_1 \right\}
\ee
where we have put $V=\sqrt{1+CR^2}$. Continuity of $g_{tt}$ settles the value of
\be
C= \frac{2M}{R^2 (R-2M)}
\ee
Now all integration constants are found and the matching is complete.

\section{Conclusion}

We have investigated whether the trace-free Einstein equations admit exact solutions that could model compact objects. The master isotropy equation was solved in conjunction with the conservation of energy momentum equation and exact solutions were found. We examined the Schwarzschild geometrical  ansatz and found that the resulting density was not constant in TFE. Imposing the requirement of constant density yielded a metric that was no longer of Schwarzschild form. The case of the isothermal sphere was considered and solved. It turned out that the constant gravitational potential which is a necessary and sufficient condition for isothermal behaviour was not the case for TFE. While the energy density obeyed the inverse square law, the equation of state requiring the pressure to be proportional to the density does not hold. Finally we considered the Finch--Skea ansatz and the model was completely determined in terms of trigonometric functions. Through the use of graphical plots, it was established that a hypersurface of vanishing pressure did exist, the density and pressure are always positive within the radius, the energy conditions are satisfied and the fluid was causal in that the speed of sound remains subluminal within the radius. Therefore this model satisfies all the essential basic requirements for physical reasonableness as a static star model. It remains to analyse other exact solutions in standard general relativity, such as the Tolman--Bondi model,  to reveal their dynamical counterparts in the trace-free scenario. This is currently under investigation.

\section*{ACKNOWLEDGEMENTS}
The authors thank the unnamed referee for drawing our attention to relevant papers such as the work of B\"{o}hmer and Winter. 


\end{document}